\documentclass[aps,prd,preprintnumbers,nofootinbib,floatfix,11pt]{revtex4}

\usepackage{amsfonts,amsmath,amssymb,mathrsfs,graphicx,epsfig,color,amsthm} 
\usepackage{hyperref}

\allowdisplaybreaks[1]
\newcommand{\be}{\begin{equation}}
\newcommand{\ee}{\end{equation}}
\newcommand{\ba}{\begin{eqnarray}}
\newcommand{\ea}{\end{eqnarray}}

\usepackage{}

\usepackage[normalem]{ulem}

\begin{document}

\title{Refined inequalities for loosely trapped surface/attractive gravity probe surface}
\author{Kangjae Lee${}^1$, Tetsuya Shiromizu$^{1,2}$ and Keisuke Izumi$^{2,1}$}

\affiliation{$^{1}$Department of Mathematics, Nagoya University, Nagoya 464-8602, Japan}
\affiliation{$^{2}$Kobayashi-Maskawa Institute, Nagoya University, Nagoya 464-8602, Japan} 

\begin{abstract}
\begin{center}
{\bf Abstract}
\end{center}
\noindent
We reexamine a loosely trapped surface (LTS) proposed as an indicator for strong gravity 
and an attractive gravity probe surface (AGPS) as that for gravity. 
Refined inequalities for them are derived by taking account of angular momentum, gravitational 
waves and matters. 
\end{abstract}

\maketitle

%
%
%
\section{Introduction}
\label{section1}

Inspired by circular orbits of photon in the Schwarzschild spacetime, a loosely trapped surface (LTS) 
has been proposed as an indicator for strong gravity~\cite{shiromizu2017}. Under certain conditions, 
one could show that the area $A_{\rm LTS}$ of an LTS satisfies (See also Ref.~\cite{hod})
\begin{equation}
A_{\rm LTS} \leq 4\pi (3m)^2, \label{lts-ineq}
\end{equation}
where $m$ is the Arnowitt-Deser-Misner (ADM) mass. The inequality for an LTS has also been examined in 
the Einstein-Maxwell system~\cite{Lee2020}. This is regarded as an analogy of the Penrose 
inequality~\cite{penrose} which has been proven for the apparent 
horizon on the time-symmetric initial data~\cite{wald, imcf, bray}. 
Suprisingly, it has been proven that an attractive gravity probe surface (AGPS), which can exist even 
in a weak gravity region, satisfies the areal inequality~\cite{Izumi2021}
\begin{equation}
A_{\rm AGPS} \leq 4\pi [(3+4\alpha)m/(1+2\alpha)]^2, \label{agps-ineq}
\end{equation}
where $\alpha >-1/2$ is a parameter appearing in the definition of AGPS. 
The lower bound of $\alpha$, $\alpha \to -1/2$, includes the case where the surface becomes 
the round sphere at spatial infinity and thus
there is no upper bound for $A_{\rm AGPS}$ in this limit. 
Meanwhile, 
an AGPS becomes a minimal surface and 
an LTS for $\alpha \to \infty$ and $\alpha=0$ respectively. A minimal surface is also an 
indicator for strong gravity, especially black holes, because an apparent horizon is a minimal surface on 
the time-symmetric initial data. 

In this paper, we shall refine the inequality for an AGPS so that the contribution from angular momentum, 
gravitational waves and matters will be taken into account. 
In the formation of a black hole through gravitational collpase, if the cosmic censorship conjecture holds and 
the system settles down to the stationary state so that spacetime will be expressed by the Kerr solution 
due to the black hole uniqueness theorem~\cite{uniq}, 
the area theorem \cite{hawking} provides us the inequality for the area of the cross section of the event 
horizon with a time slice as 
\begin{equation}
A_H \leq A_{\rm Kerr}:=8\pi m (m+{\sqrt {m^2-J^2/m^2}}), \label{kerr1}
\end{equation}
where $m$ is the black hole mass and $J$ is the angular momentum. 
This inequality can be equivalently written as
\begin{equation}
m^2 \geq \frac{A_H}{16 \pi}+4\pi \frac{J^2}{A_H}=\Bigl( \frac{{\cal R}_H}{2}\Bigr)^2+\frac{J^2}{{\cal R}_H^2},\label{kerr2}
\end{equation}
when $A_H \geq 8\pi J$ holds,%
\footnote{In axisymmetric spacetimes, under certain conditions, one can show this 
inequality~\cite{Dain2011, Acena:2010ws, Jaramillo:2011pg}} where ${\cal R}_H:={\sqrt {A_H/4 \pi}}$. 
There are many efforts to prove this~\cite{Dain2018, Anglada2018, Anglada2020}. In particular, using the inverse 
mean curvature flow and introducing a kind of radius, Anglada addressed this inequality for a minimal surface 
in axisymmetric spacelike hypersurfaces~\cite{Anglada2018}. 
We will apply Anglada's approach~\cite{Anglada2018} (See also Refs. \cite{Anglada2020, Anglada:2016dbu}) 
based on the monotonicity of Geroch's mass \cite{geroch, wald} to an AGPS, but refine the inequality for 
general cases without the axisymmetric assumption. Since an AGPS becomes a minimal surface in the limit 
$\alpha \to \infty$, our result recovers Anglada's inequality for axisymmetric cases. 

The rest of this paper is organized as follows. In Sec. \ref{sec:formula}, we will give the definition of AGPS (and LTS) 
and then present some key equations for the next two sections. In Sec. \ref{sec:general}, introducing an area-averaged 
energy density and a kind of angular momentum,  we will show the refined inequality for general cases. In Sec. \ref{sec:vacuum}, 
we will discuss the vacuum and axisymmetric cases. 
Then, since the Komar angular momentum can be employed, one can have 
more precise refined inequalities for an AGPS, an LTS and a minimal surface. Finally we will give summary and discussion in Sec. 
\ref{sec:summary}.

%
\section{Definition of LTS/AGPS and some key formulas}
\label{sec:formula}

In this paper, we show the refined inequalities for a minimal surface (MS), a loosely trapped surface (LTS) 
and an attractive gravity probe surface (AGPS). Since the first two are included into the third, we first give the 
definition of AGPS. 
An AGPS is defined as a 2-dimensional surface in a 3-dimesnional spacelike hypersurface $\Sigma$ such that it has 
the positive mean curvature $k$ and satisfies~\cite{Izumi2021}
\begin{equation}
r^aD_ak/k^2 \geq \alpha, \label{agps}
\end{equation}
where $r^a$ is the outward unit normal vector to $S_y$, $D_a$ is the covariant derivative of $\Sigma$
and $\alpha$ is a parameter satisfying $\alpha >-1/2$. In the limit of $\alpha \to \infty$, 
the surface corresponds to a minimal surface with $k=0$. For $\alpha =0$, the surface becomes an LTS~\cite{shiromizu2017}. 

Here note that, on a spacelike hypersurface $\Sigma$ with a foliation $\lbrace S_y \rbrace_{y \in {\bf R}}$,
the following equation holds  
\begin{equation}
r^aD_ak=-\varphi^{-1}{\cal D}^2\varphi-\frac{1}{2}{}^{(3)}R+\frac{1}{2}{}^{(2)}R-\frac{3}{4}k^2
-\frac{1}{2}\tilde k_{ab} \tilde k^{ab}, \label{1dk}
\end{equation}
where $\varphi$ is the lapse function so that $r^a$ is written as 
$r_a=\varphi D_a y$, ${\cal D}_a$ is the covariant derivative of $S_y$, ${}^{(3)}R$ is the 3-dimensional Ricci scalar, 
${}^{(2)}R$ is the Ricci scalar of $S_y$ and $\tilde k_{ab}$ is the traceless part of the extrinsic curvature 
$k_{ab}$ of $S_y$. 

${}^{(3)}R$ can be related to the energy density of matters, $\rho$, through the Hamiltonian constraint
\begin{eqnarray}
{}^{(3)}R+K^2-K_{ab}K^{ab}=16\pi \rho, \label{hami}
\end{eqnarray}
where $K_{ab}$ is the extrinsic curvature of $\Sigma$ and $K$ is its trace. Now we 
decompose $K_{ab}$ as 
\begin{equation}
K_{ab}=\kappa_{ab}+v_a r_b+v_br_a+K_{(r)}r_ar_b, \label{extcur}
\end{equation}
where $\kappa_{ab}:=h_a^ch_b^dK_{cd}$, $v_a:=h_a^br^cK_{bc}$, $K_{(r)}:=r^ar^bK_{ab}$ and $h_{ab}$ is the 
induced metric of $S_y$. $v_a$ is regarded as an angular velocity. 
Equations (\ref{hami}) and (\ref{extcur}) tell us that the 3-dimensional Ricci scalar is written as 
\begin{equation}
{}^{(3)}R=16 \pi \rho -\frac{1}{2}\kappa^2-2\kappa K_{(r)}+\tilde \kappa_{ab} \tilde \kappa^{ab}+2v_av^a, \label{3d-ricci}
\end{equation}
where $\kappa$ and $\tilde \kappa_{ab}$ are the trace and the traceless part of $\kappa_{ab}$ respectively.

Hereafter let us assume that $\Sigma$ is a spacelike maximal hypersurface,\footnote{
It is easy to see that the 3-dimensional Ricci scalar is non-negative when $\rho$ is non-negative 
and the above slice condition holds. 
The proof in this paper works under weaker assumptions:
(i)For $K>0$, $\kappa \leq 0$ or $\kappa \geq \frac{4}{3} K$, 
(ii)For $K<0$, $\kappa \leq \frac{4}{3}K$ or $\kappa \geq 0.$ 
They show that the sum of the second and third terms in the right-hand side of Eq. (\ref{3d-ricci}) is non-negative, 
$-\kappa^2/2-2\kappa K_{(r)} \geq 0$.} that is, $K=0$. 
With the condition (\ref{agps}) in the definition of AGPS and the maximal slice condition, 
the  surface integral of Eq. (\ref{1dk}) implies us 
\begin{equation}
\Bigl( 1+\frac{4}{3}\alpha \Bigr)\int_{S_0}dAk^2 \leq \frac{16\pi}{3}-\frac{2}{3}\int_{S_0}dA
(16\pi \rho_{\rm tot}+2v_av^a), \label{k-ineq}
\end{equation}
where $\rho_{\rm tot}:=\rho+\rho_{\rm gw}$ and 
\begin{equation}
8\pi \rho_{\rm gw}:=\frac{1}{2}(\tilde \kappa_{ab} \tilde \kappa^{ab}+\tilde k_{ab}\tilde k^{ab} ). \label{gw}
\end{equation}
Here, $\rho_{\rm gw}$ may be regarded as a part of local energy density of gravitational waves.  
Equation (\ref{k-ineq}) is the refined version of Eq. (10) in Ref.~\cite{Izumi2021} for $S_y \approx S^2$ 
and for zero cosmological constant. 

We take the inverse mean curvature flow (IMCF), that is $\varphi k=1$, and assume that the foliation 
can be taken globally.%
\footnote{It is not guaranteed that the inverse mean curvature flow can be taken globally. 
However, the singularity of the flow is resolved by Huisken and Ilmanen~\cite{imcf}, and 
the monotonicity of the Geroch mass holds true in the resolved flow. 
We expect that the similar discussion works in our cases.}
We also introduce 
the Geroch mass defined by~\cite{geroch}
\begin{equation}
E(y):=\frac{A^{1/2}(y)}{64\pi^{3/2}}\int_{S_y} \Bigl(2{}^{(2)}R-k^2 \Bigr)dA, \label{geroch}
\end{equation}
where $A(y)$ is the area of $S_y$. 
By virtue of IMCF, the first derivative of $E$ has the following simple form
\begin{eqnarray}
\frac{dE}{dy}=\frac{A^{1/2}}{64\pi^{3/2}}\int_{S_y}  \Bigl[ 2\varphi^{-2}({\cal D} \varphi)^2+{}^{(3)}R
+\tilde k_{ab} \tilde k^{ab}  \Bigr]dA. \label{yd-geroch1}
\end{eqnarray}
Using Eq. (\ref{3d-ricci}), the maximal slice condition and the assumption of $\rho \geq 0$, 
we have 
\begin{equation}
\frac{dE}{dy}
=\frac{A^{1/2}}{64 \pi^{3/2}} \int_{S_y}
\Bigl[2\varphi^{-2}({\cal D}\varphi)^2+16\pi \rho_{\rm tot}+2v_av^a   \Bigr]dA \geq 0. \label{yd-geroch2}
\end{equation}
Its integration over $y$ gives us 
\begin{eqnarray}
m_{\rm ADM}-\frac{{\cal R}_{A0}}{2}+\frac{A_0^{1/2}}{64\pi^{3/2}}\int_{S_0}dAk^2  
& \geq &  \int_0^\infty dy \frac{A^{1/2}}{64\pi^{3/2}} \int_{S_y}dA(16\pi \rho_{\rm tot}+2v_av^a) \nonumber \\
& = & 2\pi \int_0^\infty dy {\cal R}_A^3 \bar \rho_{\rm tot}+\frac{1}{16\pi}\int_0^\infty dy {\cal R}_A \int_{S_y} dAv^av_a 
\nonumber \\
& = & m_{\rm ext}+\frac{1}{16\pi}\int_0^\infty dy {\cal R}_A \int_{S_y} dAv^av_a,
\label{E-integral}
\end{eqnarray}
where we used the fact that $E(\infty)=m_{\rm ADM}$, and the Gauss-Bonnet theorem $\int_{S_0}dA{}^{(2)}R=8 \pi$ for the left-hand side.
Hereinafter, variables with subscript $0$, such as $A_0$, are those evaluated on $S_0$, which corresponds to the surface for $y=0$. 
In the second line of Eq. (\ref{E-integral}), we used the area radius defined by ${\cal R}_A(y):={\sqrt {A/4\pi}}$ and 
the surface-averaged energy density 
\begin{equation}
\bar \rho_{\rm tot}(y):=\frac{1}{A}\int_{S_y}dA \rho_{\rm tot}. \label{averaged-rho}
\end{equation}
In the third line, we put 
\begin{eqnarray}
m_{\rm ext}:=2\pi \int_0^\infty dy {\cal R}_A^3 \bar \rho_{\rm tot}, \label{ext-mass}
\end{eqnarray}
which is the total rest mass of the matters and gravitational waves in the region between $y=0$ and infinity. 
Since the fact that ${\cal R}_{A} \propto e^{y/2}$ in IMCF gives $2\pi \int dy {\cal R}_A^3=(4\pi/3){\cal R}_A^3$, 
definition (\ref{ext-mass}) is merely natural. In the following sections, $S_0$ will be supposed to be an MS/LTS/AGPS.

%
\section{Refined inequalities for general cases}
\label{sec:general}

With the introduction of a few quantities, Eqs. (\ref{k-ineq}) and (\ref{E-integral}) are summarized as a theorem:

\noindent 
{\it Theorem 1:~
Let $\Sigma$ be an asymptotically flat spacelike maximal hypersurface having the inverse mean curvature 
flow $\lbrace S_y \rbrace_{y \in {\bf R}}$ with $S_y \approx S^2$. Assuming that the energy density 
$\rho$ appearing in the Hamiltonian constraint is non-negative, then, we have an inequality 
for an AGPS
\begin{eqnarray}
m_{\rm ADM}-\Bigl(m_{\rm ext}+\frac{3}{3+4\alpha}m_{\rm int}  \Bigr) 
& \geq & \frac{1+2\alpha}{3+4\alpha} {\cal R}_{A0}+\frac{1}{{\cal R}^3_{A0}} 
\Bigl( \frac{3}{3+4\alpha}\bar J_0^2+\bar J_{\rm min}^2 \Bigr) \label{pi-ineq} \\ 
& \geq & \frac{1+2\alpha}{3+4\alpha} {\cal R}_{A0}+2\frac{3+2\alpha}{3+4\alpha}\frac{\bar J_{\rm min}^2 }{{\cal R}^3_{A0}},
\label{pi-ineq1}
\end{eqnarray}
where 
\begin{eqnarray}
m_{\rm int}:=\frac{4\pi}{3}{\cal R}_{A0}^3 \bar \rho_{{\rm tot}0}, \label{mass-int}
\end{eqnarray}
\begin{eqnarray}
\Bigl(8\pi \bar J (y)\Bigr)^2:=\frac{A^2}{6\pi} \int_{S_y}v_av^adA, \label{ave-j}
\end{eqnarray}
and
\begin{eqnarray}
\bar J_{\rm min}:=\min_{\lbrace S_y \rbrace}\bar J. \label{min-j}
\end{eqnarray}
}
\begin{proof}
Using $\bar J$ defined by Eq. (\ref{ave-j}), we write Eq. (\ref{k-ineq}) as
\begin{eqnarray}
\Bigl(1+\frac{4}{3}\alpha \Bigr)\int_{S_0}dAk^2 \leq \frac{16\pi}{3}-\frac{32\pi}{3}A_0 
\bar \rho_{{\rm tot}0}-32 \pi\frac{\bar J_0^2}{{\cal R}^4_{A0}}. \label{k-ineq2}
\end{eqnarray}
For the last term in the right-hand side of Eq. (\ref{E-integral}), we can see 
\begin{eqnarray}
\frac{1}{16\pi}\int_0^\infty dy {\cal R}_A \int_{S_y} dAv^av_a
=\frac{3}{2}\int_0^\infty dy \frac{\bar J^2}{{\cal R}^3_A} \geq \frac{3}{2}{\bar J}_{\rm min}^2 \int_0^\infty \frac{dy}{{\cal R}^3_A}
=\frac{\bar J^2_{\rm min}}{{\cal R}^3_{A0}}.
\end{eqnarray}
Then, after simple manipulation, it is easy to see that Eqs. (\ref{E-integral}) and (\ref{k-ineq2}) 
imply us Eqs. (\ref{pi-ineq}) and (\ref{pi-ineq1}). 
\end{proof}

We can see that inequality (\ref{pi-ineq}) or (\ref{pi-ineq1}) is the refined version of  Eq. (\ref{agps-ineq}) obtained for an AGPS in Ref.~\cite{Izumi2021}. 
The inequality is satisfied even if the non-negative quantities $m_{\rm ext}$, $m_{\rm int} $ and $\bar J_{\rm min}^2$ are set to be zero in Eq. (\ref{pi-ineq1}), 
and Eq. (\ref{agps-ineq}) is recovered.

We have some remarks on the definitions introduced for the theorem 1: (i)\,$m_{\rm int}$ defined 
by Eq. (\ref{mass-int}) may be regarded as a mass in the region surrounded by $S_0$. (ii)\,The definition of 
the area-averaged angular momentum (\ref{ave-j}) comes from the observation for spherically symmetric 
cases and asymptotic behavior. In this sense, the validity of the definition for general cases is 
far from canonical one based on conservation. In the next section, nevertheless, we show the magnitude 
relation between the area-averaged angular momentum $\bar J$ and
the Komar angular momentum for vacuum and axisymmetric cases. 

For an LTS ($\alpha=0$), Eqs. (\ref{pi-ineq}) and (\ref{pi-ineq1}) become 
\begin{equation}
m_{\rm ADM}-(m_{\rm ext}+m_{\rm int}) 
\geq \frac{{\cal R}_{A0}}{3}+\frac{\bar J_0^2+\bar J^2_{\rm min}}{{\cal R}^3_{A0}} \geq \frac{{\cal R}_{A0}}{3}+2\frac{\bar J^2_{\rm min}}{{\cal R}^3_{A0}}.
\end{equation}
This includes Eq. (\ref{lts-ineq}) obtained in Ref.~\cite{shiromizu2017}. 
For a minimal sufrace ($\alpha \to \infty$), Eq. (\ref{pi-ineq1}) becomes 
\begin{eqnarray}
m_{\rm ADM}-m_{\rm ext} \geq \frac{{\cal R}_{A0}}{2}+\frac{\bar J^2_{\rm min}}{{\cal R}^3_{A0}}.
\label{mADM-mext}
\end{eqnarray}
From this, one can obtain the Penrose inequality, $A_0 \leq 4\pi(2m_{\rm ADM})^2$, shown in Ref.~\cite{wald} 
(See also Refs.~\cite{imcf, bray}). 
We can see the similarity to Eq. (\ref{kerr2}) by taking the square of Eq. (\ref{mADM-mext}),
\begin{eqnarray}
(m_{\rm ADM}-m_{\rm ext})^2 \geq \Bigl(\frac{{\cal R}_{A0}}{2}\Bigr)^2+\frac{\bar J^2_{\rm min}}{{\cal R}^2_{A0}}
+ \Bigl( \frac{\bar J^2_{\rm min}}{{\cal R}^3_{A0}} \Bigr)^2 \geq \Bigl(\frac{{\cal R}_{A0}}{2}\Bigr)^2+\frac{\bar J^2_{\rm min}}{{\cal R}^2_{A0}}. 
\end{eqnarray}
Note that an AGPS with $\alpha$ being close to minus one half  can exist near asymptotic infinity and thus our inequality for an AGPS holds even for weak gravity. 

The arithmetic-geometric mean of the right-hand side of Eq. (\ref{pi-ineq1}) gives us a Corollary:

\noindent 
{\it Corollary~2:~
In the same setup and assumption of Theorem 1, 
\begin{equation}
\Delta m_{\rm ADM} \geq 2 C_\alpha \frac{\bar J_{\rm min}}{{\cal R}_{A0}}
\end{equation}
holds for AGPS,  where 
\begin{equation}
\Delta m_{\rm ADM}:=m_{\rm ADM}-\Bigl(m_{\rm ext}+\frac{3}{3+4\alpha}m_{\rm int}  \Bigr) 
\end{equation}
and
\begin{equation}
C_\alpha:={\sqrt {\frac{2(1+2\alpha)(3+2\alpha)}{(3+4\alpha)^2}}}.
\end{equation}
}

This corollary gives the lower bound for ${\cal R}_{A0}$ as 
\begin{equation}
{\cal R}_{A0} \geq 2C_\alpha \frac{\bar J_{\rm  min}}{\Delta m_{\rm ADM}}. 
\label{R>J}
\end{equation}
It is interesting to compare to the universal inequality ${\cal R} \gtrsim J^{1/2} $ for axisymmetric 
rotating body shown by Dain~\cite{Dain2014} (See also Refs.~\cite{Khuri:2015xpa, Reiris:2014tva}). 
The ratio $\epsilon_\alpha$ of the lower bound for ${\cal R}_{A0}$ to Dain's one is 
\begin{equation}
\epsilon_\alpha \sim \frac{J/m}{J^{1/2}} \sim a^{1/2},
\end{equation}
where $a:=J/m^2$ is the Kerr parameter. 
Note that an AGPS can exist in a weak gravity region and does not require a black hole,
that is, inequality (\ref{R>J}) can be applied not only to black hole but also to other objects such as a star. 
For astrophysical objects except for compact objects, 
$a$ is much larger than unity. 
Therefore, our inequality is relatively strong for such cases.

In the limit of $\alpha=-1/2$, $C_{-1/2}$ vanishes and thus inequality (\ref{R>J}) does not give any constraint.
Going back to the original inequality (\ref{pi-ineq1}), however, we can give another lower bound on ${\cal R}_{A0}$.
Since the first term in the right-hand side of inequality (\ref{pi-ineq1}) is non-negative for $\alpha\ge -1/2$, 
we have a weaker inequality,
\begin{equation}
m_{\rm ADM}-\left(m_{\rm ext}+ \frac{3}{3+4\alpha}m_{\rm int} \right) \geq 2\frac{3+2\alpha}{3+4\alpha}\frac{\bar J^2_{\rm min}}{{\cal R}^3_{A0}}.
\end{equation}
Due to the fact that $m_{\rm ext} \geq 0$, $m_{\rm int} \geq 0$, we have
\begin{equation}
m_{\rm ADM}\geq 2\frac{3+2\alpha}{3+4\alpha}\frac{\bar J^2_{\rm min}}{{\cal R}^3_{A0}}.
\end{equation}
This is rearranged to 
\begin{equation}
{\cal R}_{A0} \geq \Bigl( 2\frac{3+2\alpha}{3+4\alpha} \frac{\bar J^2_{\rm min}}{m_{\rm ADM}} \Bigr)^{1/3}.
\end{equation}
Unlike inequality (\ref{R>J}), this inequality gives a meaningful condition for $\alpha=-1/2$.
We could have the lower bound for the area radius of AGPS with $\alpha=-1/2$ and $\epsilon_{-1/2}\sim a^{1/6}$.

%

\section{Vacuum and axisymmetric cases}
\label{sec:vacuum}

In this section, we consider vacuum and axisymmetric cases. Let $\phi^a$ be the axisymmetric Killing vector. 
Then, we can define the Komar angular momentum $J$~\cite{Komar:1958wp} by 
\begin{equation}
J:=\frac{1}{8\pi}\int_{S_y}v^a \phi_a dA. \label{komar-J}
\end{equation}
It is easy to see that $J$ does not depend on $y$ due to the vacuum. 

In terms of the conserved angular momentum $J$, one can refine the contribution from the angular momentum for the 
inequality in the previous section as below: 

\noindent 
{\it Theorem 3:~
In vacuum and axisymmetric spacetimes,  let $\Sigma$ be an asymptotically flat  axisymmetric spacelike maximal hypersurface 
having the inverse mean curvature flow $\lbrace S_y \rbrace_{y \in {\bf R}}$ 
with $S_y \approx S^2$. Assuming that the energy density $\rho$ is non-negative, then, we have an inequality for an AGPS
\begin{equation}
m_{\rm ADM} -\Bigl(m_{\rm gw,ext}+\frac{3}{3+4\alpha}m_{\rm gw,int}  \Bigr) 
\geq \frac{1+2\alpha}{3+4\alpha}{\cal R}_{A0}+\frac{1+\gamma_\alpha}{{\cal R}_0^2{\cal R}_{A0}}J^2,\label{pi-ineq2}
\end{equation}
where 
\begin{equation}
\frac{1}{{\cal R}_0^2}:=\frac{3}{2}{\cal R}_{A0}\int^\infty_0\frac{{\cal R}_A}{{\cal R}^4_\phi}dy,
\end{equation}
\begin{equation}
\gamma_\alpha:=\frac{3}{3+4\alpha}\frac{{\cal R}_0^2{\cal R}_{A0}^2}{{\cal R}_{\phi 0}^4},
\label{gammaalpha}
\end{equation}
\begin{equation}
\frac{8\pi}{3}{\cal R}_\phi^4 (y) :=\int_{S_y}\phi_a \phi^a dA,
\end{equation}
\begin{eqnarray}
m_{\rm gw,ext}:=2\pi \int_0^\infty dy {\cal R}_A^3 \bar \rho_{\rm gw}
\end{eqnarray}
and
\begin{eqnarray}
m_{\rm gw,int}:=\frac{4\pi}{3}{\cal R}_{A0}^3 \bar \rho_{{\rm gw}0}. 
\end{eqnarray}
}
\begin{proof}
Note that, using the Cauchy-Schwarz inequality and the definition of $J$, one can show
\begin{equation}
\int_{S_y}v^av_adA  \int_{S_y} \phi^a \phi_a dA \geq \Bigl(\int_{S_y}v^a \phi_a dA \Bigr)^2=(8\pi J)^2. \label{v^2}
\end{equation}
Then, Eq. (\ref{k-ineq}) is expressed with ${\cal R}_{\phi0}$ as  
\begin{equation}
\Bigl(1+\frac{4}{3}\alpha \Bigr)\int_{S_0}dAk^2 \leq \frac{16\pi}{3}-\frac{32\pi}{3}A_0\bar \rho_{{\rm gw}0}
-32\pi \frac{J^2}{{\cal R}_{\phi 0}^4}. \label{k-ineq3}
\end{equation}
Inequality (\ref{v^2}) gives a lower bound for the second term of the right-hand side of Eq. (\ref{E-integral}), 
\begin{eqnarray}
\frac{1}{16\pi}\int_0^\infty dy {\cal R}_A \int_{S_y} dAv^av_a \geq \frac{J^2}{{\cal R}_0^2{\cal R}_{A0}}.
\end{eqnarray}
Finally, Eq. (\ref{E-integral}) with Eq. (\ref{k-ineq3}) presents Eq. (\ref{pi-ineq2}). 
\end{proof}

In addition to ${\cal R}$ introduced in Ref.~\cite{Anglada2018}, 
we defined here the new radius ${\cal R}_\phi$. For spherically symmetric cases, 
both of them coinside with the area radius ${\cal R}_A$, that is, ${\cal R}={\cal R}_\phi={\cal R}_A$. 
Furthermore, for a convex $S_y$, we can show (See Appendix A for the details) 
\begin{equation}
\frac{1}{3} \leq \frac{{\cal R}_{A}^2{\cal R}^2}{{\cal R}_\phi^4} \leq \frac{5}{3}. \label{radius-relation}
\end{equation}
This constrains $\gamma_\alpha$ as 
\begin{equation}
\frac{1}{3+4\alpha} \leq \gamma_\alpha \leq \frac{5}{3+4\alpha}. \label{gamma-alpha}
\end{equation}
One can also show that for $\lambda_\theta \geq \lambda_\phi >0$ (oblate case) 
\begin{equation}
{\cal R}_\phi \geq {\cal R}_A \label{radius-relation2}
\end{equation}
holds, while for $0<\lambda_\theta \leq \lambda_\phi$ (prolate case)
\begin{equation}
{\cal R}_\phi \leq {\cal R}_A \label{radius-relation3}
\end{equation}
holds, where $\lambda_\phi$ and  $\lambda_\theta$  are the principal curvatures of $S_y$ with respect to the Killing direction $\phi^a$ and that normal to $\phi^a$, respectively,   so that $k=\lambda_\theta+\lambda_\phi$ (See Appendix A for the details). 

From the definition (\ref{ave-j}) and the Cauchy-Schwarz inequality (\ref{v^2}), it is easy to see that 
the relation between $\bar J$ and $J$ 
\begin{equation}
\bar J^2 \geq \Bigl(\frac{{\cal R}_A}{{\cal R}_\phi} \Bigr)^4J^2
\end{equation}
holds. For $0<\lambda_\theta \leq \lambda_\phi$, together with Eq. (\ref{radius-relation3}), it tells us 
\begin{equation}
\bar J^2 \geq J^2. 
\end{equation}

For an LTS ($\alpha=0$), Eq. (\ref{pi-ineq2}) becomes 
\begin{equation}
m_{\rm ADM} -(m_{\rm gw, ext}+m_{\rm gw,int})
\geq \frac{{\cal R}_{A0}}{3}+(1+\gamma_0) \frac{J^2}{{\cal R}_0^2{\cal R}_{A0}}.
\end{equation}
For a minimal surface ($\alpha \to \infty$), we recover Anglada's result~\cite{Anglada2018}
\begin{equation}
m_{\rm ADM} -m_{\rm gw,ext}
\geq \frac{{\cal R}_{A0}}{2}+\frac{J^2}{{\cal R}_0^2{\cal R}_{A0}}.
\end{equation}
Taking of the square of this gives a similar inequality to Eq. (\ref{kerr2}),
\begin{equation}
m_{\rm ADM}^2\geq \Bigl( \frac{{\cal R}_{A0}}{2} \Bigr)^2+\frac{J^2}{{\cal R}_0^2}
+\Bigl( \frac{J^2}{{\cal R}_0^2{\cal R}_{A0}} \Bigr)^2 \geq \Bigl( \frac{{\cal R}_{A0}}{2} \Bigr)^2+\frac{J^2}{{\cal R}_0^2}.
\end{equation}

Applying the arithmetic-geometric mean for Eq. (\ref{pi-ineq2}), we also have a similar result 
to Corollary 2: 

\noindent 
{\it Corollary 4:~
In the same setup and assumption of Theorem 3, 
\begin{equation}
m_{\rm ADM} -\Bigl(m_{\rm gw,ext}+\frac{3}{3+4\alpha}m_{\rm gw,int}  \Bigr) 
\geq 2F_\alpha  \frac{|J|}{{\cal R}_0}
\label{Cor4}
\end{equation}
holds for an AGPS, where 
\begin{equation}
F_\alpha:={\sqrt {\frac{(1+\gamma_\alpha)(1+2\alpha)}{3+4\alpha}}}. 
\end{equation}
}

Note that $F_\alpha$ depends on radii (See the definition of $\gamma_\alpha$, Eq. (\ref{gammaalpha})). 
For convex $S_y$, however, Eq. (\ref{gamma-alpha}) for $\gamma_\alpha$ 
gives a lower bound of $F_\alpha$ as $F_\alpha \geq 2{\sqrt {(1+\alpha)(1+2\alpha)/(3+4\alpha)^2}}=:F_{\rm min}$. 
Since $F_{\rm min}$ is independent of radii, Eq. (\ref{Cor4}) gives a lower bound for ${\cal R}_0$ of AGPS,  
\begin{equation}
{\cal R}_0 \geq 2F_{\rm min} \frac{|J|}{m_{\rm ADM}}.
\end{equation}
In the limit of $\alpha=-1/2$, one can have a similar lower bound for a combination of radii, but 
its form is not simple. Since the argument based on the order of magnitude is the same with that in the 
previous section, here we do not show the derivation again.


\section{Summary and discussion}
\label{sec:summary}

In this paper, taking account of the contributions from angular momentum, 
gravitational waves and matters, we have shown the refined inequalities for an attravtive 
gravity probe surface (AGPS) which includes a minimal surface (MS) and a loosely trapped surface (LTS). 
We have also discussed the relation among newly introduced radii. 

In order to obtain the inequalities in a sophisticated form,  new quasi-local quantities 
related to the size, rotation and gravitational wave are required to be introduced. 
The physical meaning of some of them are not totally clear. 
For example, we have defined the energy density for gravitational waves. 
It is similar to that obtained through the argument for the linear perturbation of metric, but not exactly the same. 
This may indicate a possiblity of an improvement of our argument for MS/LTS/AGPS. 

\acknowledgments

T. S. and K. I.  are supported by Grant-Aid for Scientific Research from Ministry of Education, 
Science, Sports and Culture of Japan (Nos. 17H01091, JP21H05182, JP21H05189). 
T. S. is also supported by JSPS Grants-in-Aid for Scientific Research (C) (JP21K03551). 
K.~I. is also supported by JSPS Grants-in-Aid for Scientific Research (B) (JP20H01902).

\appendix

%
%
\section{Relation between ${\cal R}$, ${\cal R}_A$ and ${\cal R}_\phi$}
\label{sec:appendix}

In this appendix, following Ref.~\cite{Anglada2018}, we show Eq. (\ref{radius-relation}). 
In IMCF, using $A \propto e^y$, we have  
\begin{eqnarray}
\frac{d}{dy}\Biggl( \frac{A^{1/2}}{\int_{S_y}\phi_a \phi^adA} \Biggr)& = & -\frac{1}{2}\frac{A^{1/2}}{\int_{S_y}\phi_a \phi^adA}
-2A^{1/2}\frac{ \int_{S_y}\frac{\lambda_\phi}{k}\phi_a \phi^a dA}{\Bigl( \int_{S_y}\phi_a \phi^adA \Bigr)^2} \label{a1} \\
& = & -\frac{5}{2}\frac{A^{1/2}}{\int_{S_y}\phi_a \phi^adA}
+2A^{1/2}\frac{ \int_{S_y}\frac{\lambda_\theta}{k}\phi_a \phi^a dA}{\Bigl( \int_{S_y}\phi_a \phi^adA \Bigr)^2}, \label{a2}
\end{eqnarray}
where we used 
the fact that $\partial_y (\phi_a \phi^a)=2\varphi \lambda_\phi \phi_a \phi^a=2(\lambda_\phi/k)\phi_a \phi^a$~\cite{Anglada:2016dbu}. 

For a convex $S_y$, that is, $\lambda_\theta >0, \lambda_\phi>0$, Eqs. (\ref{a1}) and (\ref{a2}) lead us 
\begin{eqnarray}
-\frac{5}{2}\frac{{\cal R}_A}{{\cal R}_\phi^4} 
\leq  
\frac{d}{dy}\Biggl( \frac{{\cal R}_A}{{\cal R}_\phi^4}  \Biggr) 
\leq
-\frac{1}{2}\frac{{\cal R}_A}{{\cal R}_\phi^4} 
\label{a3} 
\end{eqnarray}
and then its integration over $y$ gives us  Eq. (\ref{radius-relation}). 

In a way similar to the derivation for Eq. (\ref{a2}), we see that
\begin{eqnarray}
\frac{d}{dy}\Biggl( \frac{A^2}{\int_{S_y}\phi_a \phi^adA} \Biggr)
=A^2 \frac{ \int_{S_y}\frac{\lambda_\theta -\lambda_\phi}{k}\phi_a \phi^a dA}{\Bigl( \int_{S_y}\phi_a \phi^adA \Bigr)^2}  \label{4}
\end{eqnarray}
holds. For $\lambda_\theta \geq \lambda_\phi >0$ (oblate), its integration over $y$ gives us 
\begin{eqnarray}
 \frac{{\cal R}_A}{{\cal R}_\phi} \leq   \left. \frac{{\cal R}_{A }}{{\cal R}_{\phi }} \right|_{y\to\infty} =1 ,
\end{eqnarray}
where we used ${\cal R}_A /{\cal R}_\phi \to 1$ at spatial infinity. This is reasonable because 
of the oblate shape. On the other hand, for $0<\lambda_\theta \leq \lambda_\phi$ (prolate), we have
\begin{eqnarray}
1 \leq \frac{{\cal R}_A}{{\cal R}_\phi}. 
\end{eqnarray}


\end{document}